\documentclass[rmp,aps,groupedaddress,floatfix]{revtex4}

\usepackage{graphicx}
\usepackage{color}
\usepackage{amsmath}

\newcommand{\RT}{($25^o C$)}

\renewcommand{\[}{\begin{equation}}
\renewcommand{\]}{\end{equation}}

\newcommand{\e}[1]{\ensuremath{\mathcal{E}(#1)}}
\newcommand{\kB}{\ensuremath{k_B}}
\newcommand{\mm}{\ensuremath{\mu m}}
\newcommand{\lp}{lambda phage}
\newcommand{\bphi}{BP-$\phi$X174}
\newcommand{\FEL}{FEL}
\newcommand{\ts}[1]{\ensuremath{\tau_\mathrm{#1}}}

\begin{document}

\title{Pause Point Spectra in DNA Constant-Force Unzipping}

\author{J. D. Weeks$^*$, J. B. Lucks$^\#$,
Y. Kafri$^*$, C. Danilowicz$^*$, D. R. Nelson$^*$ and M. Prentiss$^*$}
\affiliation{$^*$ Department of Physics, Harvard University,
Cambridge, MA 02138} 
\affiliation{$^\#$ Department of Chemistry and Chemical Biology,
Harvard University, Cambridge MA 02138}

\date{April 19, 2004}

\begin{abstract}
Under constant applied force, the separation of double-stranded DNA into
two single strands is known to proceed through a series of pauses and
jumps.  Given experimental traces of constant-force
unzipping, we present a method whereby the locations of pause points can
be extracted in the form of a pause point spectrum.  A simple theoretical
model of DNA constant-force unzipping is demonstrated to produce good agreement
with the experimental pause point spectrum of \lp{} DNA.  The locations
of peaks in the experimental and theoretical pause point spectra are found
to be nearly coincident below 6000 bp.
The model only requires the sequence, temperature and
a set of empirical base pair binding and stacking energy parameters, and
the good agreement with experiment suggests that pause points
are primarily determined by the DNA sequence. The model is also used to
predict pause point spectra for the BacterioPhage PhiX174 genome.
The algorithm for extracting the pause point spectrum might also be
useful for studying related systems which exhibit pausing behavior
such as molecular motors.
\end{abstract}

\maketitle

\section{Introduction}

The unbinding of double-stranded DNA (dsDNA) into single-stranded DNA
(ssDNA) is a ubiquitous event central to many cellular processes.
Much research has focused on understanding the
thermal unbinding of dsDNA \cite{Wartell85}.
These studies have revealed quantitative aspects of the
thermal unbinding transition through the extraction of sequence-dependent
free energy differences between bound and unbound DNA
\cite{SantaLucia96,Blossey03,Rouzina99a,Rouzina99b}.
In living cells, however, the unbinding
of dsDNA is typically achieved using molecular motors which
utilize chemical energy and exert forces to pull apart the
strands of dsDNA. 
To have a quantitative understanding of these processes it
is important to first study the simpler case of unbinding
of dsDNA by a \emph{constant} external force.  This process is typically
referred to as `unzipping' of dsDNA. For early experiments which
unzip lambda phage DNA with a constant velocity and a 
\emph{fluctuating} force, see \cite{Bockelmann02}.

\begin{figure}
\includegraphics[scale=0.7]{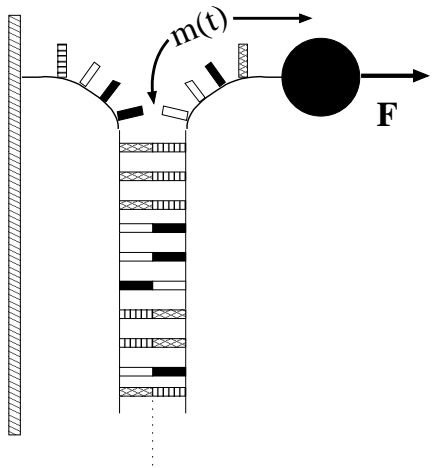}
\caption{\label{fig:model} Schematic diagram of the DNA constant-force unzipping experiment.
One strand of the dsDNA is attached to a fixed support, typically via a linker
DNA strand (not shown).  The other strand is pulled with a constant force, $F$,
via a magnetic bead (not to scale) attached to the strand.  If the force is large enough,
the dsDNA separates into two ssDNA strands.  The position of this separation,
measured in base pairs opened
$m(t)$, locates the unzipping fork. See Figure \ref{fig:exp_setup} for
a more detailed description of the experimental setup used in this paper.}
\end{figure}

Recently, single-molecule experiments have allowed study of this
process (see Fig.
\ref{fig:model} for a schematic illustration of the experiment).
Both theory \cite{Nelson03,Lubensky00,Bhattacharjee00,Sebastian00,
Cocco01,Cocco02,Marenduzzo02,Lubensky02,Kafri02}
and experiments \cite{Danilowicz03b} show that at a
given temperature the dsDNA separates into ssDNA when the applied
force exceeds a critical
value $F_c$. Moreover, for forces near $F_c$, the dynamics of the unzipping
process is highly irregular \cite{Danilowicz03a}, as displayed in
the time evolution of the junction between
the separated ssDNA and the bound dsDNA. 
This junction is referred to as the unzipping fork.
Rather than a smooth time evolution, the position of the unzipping
fork progresses through a series of long pauses separated by rapid
bursts of unzipping.

Pauses and jumps in constant-force unzipping can have several
origins.  For temperatures near the dsDNA melting transition,
portions of the dsDNA can unbind and form transient `bubbles'
below the unzipping fork.  In addition, because of
the helical nature of the dsDNA structure, a natural twist 
can be accumulated during the force-induced unzipping process.
If the unzipping fork encounters a thermal bubble in the
course of its progress, then we would expect a jump
in its position.  Furthermore,
if the unzipping occurs on time scales much faster than
the time scales associated with untwisting, then one would expect 
pauses when the DNA has to unravel accumulated 
twist \cite{Thomen02}.  Moreover, since AT and CG base pairs (bp) have different
interaction strengths (and associated base pair stacking energies
- see Table \ref{tab:bq}), pauses and jumps could also be due to
effects associated with the particular sequence of the DNA.

Experiments on multiple identical copies of the same DNA have shown
that locations of pauses are highly conserved from one strand to
another, \cite{Danilowicz03a}.  Hence, it seems likely that in these
experiments at least, transient bubbles and accumulated twist
play only a minor role in determining the jumps and pauses.
In this work we study the location of the pause points both
experimentally and theoretically.  To facilitate this study, we
introduce a pause point spectrum which is a function of the
number of base pairs unzipped.  The locations of peaks in the
pause point spectrum signify the location of pause points
in unzipping, and peak areas can be used as a measure of the
strength of pause points.  We predict pause point locations
by adapting a model of the dynamics of the unzipping fork in 
a constant-force unzipping experiment on heterogeneous DNA
\cite{Lubensky02}.
The only input information into the analysis
is the DNA sequence, free-energy differences between dsDNA and
ssDNA obtained using melting experiments, and temperature. Both
thermal bubbles and build up of twist are ignored within our
treatment. We find that we can predict most experimentally
observed pause points, thus confirming that pause point
locations are primarily a function of sequence. Our algorithm
might also prove useful for analyzing pause points arising in other
single molecule experiments, such as RNA polymerase and exonuclease
\cite{Davenport00,Wang98,Neuman03,Perkins03}.

The paper is organized as follows: In section \ref{sec:exp},
we describe constant-force unzipping experiments performed
on \lp{} DNA. In section \ref{sec:pause}, we present
 an algorithm for constructing a
pause point spectrum from experimental traces of unzipping
fork position versus time.  Section \ref{sec:theory:FEL} describes
a theoretical model of DNA constant-force unzipping which
defines a free energy landscape  as a function of
the number of bases unzipped, $m$, used to describe the
unzipping process.  In section \ref{sec:theory:MC}, this 
free energy landscape
is used as a surface on which to perform Monte Carlo 
simulations mimicking the unzipping experiments. In the same
way as performed for experimental unzipping trajectories,
these trajectories are combined to form theoretical pause
point spectra, which are compared with experiment in 
section \ref{sec:disc}.

\section{Experimental Method}
\label{sec:exp}

\begin{figure}
\includegraphics[scale=0.8]{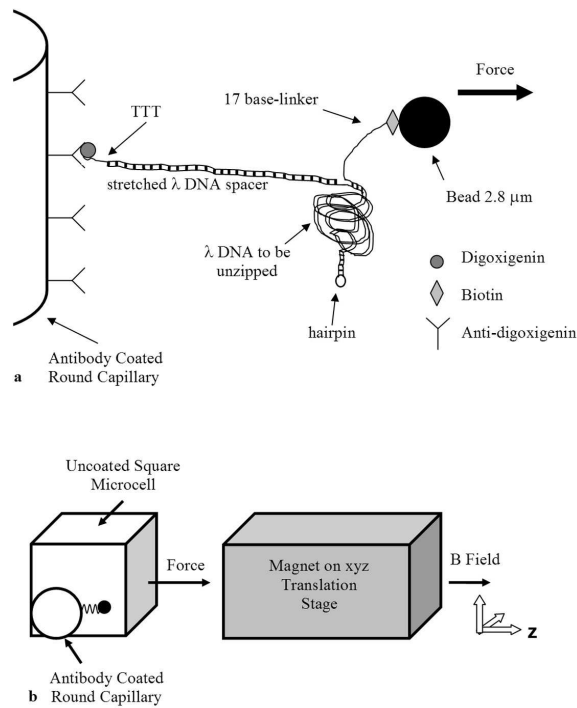}
\caption{\label{fig:exp_setup} Molecular construction and square cell.
(A) Schematic of the DNA binding to the inner glass capillary and the
magnetic bead such that pulling the bead away from the surface will cause
the dsDNA shown on the right side of the diagram to be separated into two
single DNA strands. Note that the figure is not to scale, considering that
lambda  DNA contains 48,502 bp. (B) Schematic of the side view of the
square capillary containing the round glass capillary to which the DNA
molecules are bound. The magnetic tweezer apparatus exerts the controlled
force on the magnetic beads, a microscope is used for observation, and two
thermoelectric coolers are used to control the temperature of the sample during
the initial incubation. The magnetic beads are pulled to the right in a
direction parallel to the bottom and top surfaces of the square capillary, and
perpendicular to the surface of the round capillary at a height equal to the
radius of the round capillary, where we focus the microscope. This design allows
us to view DNA molecules that are offset from the surfaces of the square capillary,
and to infer the number of separated base pairs (bp) by measuring the separation between
the magnetic bead and the surface of the round capillary.}
\end{figure}

The experimental procedure has been discussed previously in
\cite{Danilowicz03a,Assi02}.  As shown in Figure \ref{fig:exp_setup},
our setup consisted of two pieces of \lp{} DNA, covalently
bound to each other.  One strand of DNA was used as a
spacer between the glass capillary and the other strand,
which was to be unzipped.  The spacer strand of DNA was attached
to the capillary with a digoxigenin/anti-digoxigenin antibody bond.  The
capillary was coated with digoxigenin antibody while the spacer strand
of DNA was hybridized with a digoxigenin labeled oligonucleotide.  One
end of the strand of DNA to be unzipped was hybridized and ligated with a
hairpin to prevent the complete separation of the unzipped DNA
molecule.  The other end of the strand was hybridized and ligated with a
biotinylated oligonucleotide which specifically bound to a streptavidin-coated
super-paramagnetic bead.  When a magnetic field was applied, the
force induced on the bead slowly unzipped the DNA.
The unzipping direction (order of nucleotides unzipped)
was controlled by the selection of oligonucleotides.

Figure \ref{fig:exp_setup}b shows the round, antibody coated
capillary inside an uncoated square microcell.  The round capillary
was 0.5 mm in diameter, while the square microcell was 0.8 mm
across, leaving a space for a solution of DNA, beads and buffer
inside the microcell but outside the sealed, empty, round capillary.
The capillary was incubated in a solution containing digoxigenin antibody
at 5$^o$C for at least two days.  The DNA solution and bead suspension
 were also individually
kept at 5$^o$C  prior to the experiment.  We inserted a digoxigenin antibody
coated capillary and the DNA and bead  suspension into the microcell and then
incubated it at 37$^o$C for 45 minutes, allowing the DNA to bind to the
capillary via the antigen-antibody bond.  Finally, we rotated
the microcell so the beads that had settled on top of the round capillary
were hanging off of its side.  We focused using a microscope objective
on the beads that were
attached to the outermost point on the capillary so that we could
accurately measure the distance between the beads in our field of view and the capillary.

We applied a magnetic force by bringing a stack of small magnets mounted
on an xyz translation stage near the microcell.  The magnets could be
approximated as a solenoid with its long axis in the z-direction, so the
field gradient acted in the z-direction only and was essentially 
uniform \cite{Assi02} over our
field of view, which was much smaller than the solenoid radius.

\begin{figure}
\hspace{3cm}
\includegraphics[clip=true,viewport= 0cm 0cm 9cm 9cm]{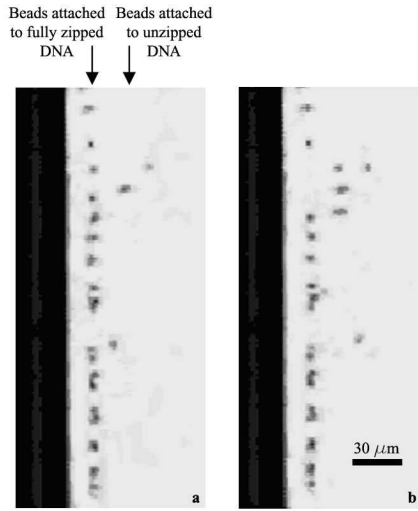}
\caption{\label{fig:exp_figs}Photographs from above the square capillary shown in 
Figure \ref{fig:exp_setup}(b).
The dark bar on the left is the inner round capillary to which 
the beads are tethered.  The dark dots are the beads.  (a)
shows the beads immediately after a 15 pN force was introduced.  Most 
of the beads are in the fully zipped position, approximately 17 \mm{} 
from the surface of the capillary.  (b) shows the beads three 
minutes later.  More DNA strands have unzipped, causing the beads to 
jump farther from the round capillary.}
\end{figure}

We measured the distance the DNA molecules had unzipped by tracking their
attached beads.  We took still digital photographs of the field of view
through a 10x objective lens once every 10 seconds.  An image processing
program found the coordinates of each bead in each frame.  
Figure \ref{fig:exp_figs}
shows part of our field of view at two different times.  
In Figure \ref{fig:exp_figs}(a), we had just applied the magnetic field.  
Figure \ref{fig:exp_figs}(b)
shows the same beads 28 frames, or just over 3 minutes later.

In each experiment, we saw approximately 50 beads in our field of view.
Approximately 10 beads unzipped slowly over the course of the experiment,
pausing at various points.  
An individual bead  paused at fixed extension until, through thermal fluctuations,
the unzipping proceeded.  After overcoming an energy barrier which we
attribute to sequence heterogeneity, the
strand unzipped up to the next pause point, where the same process repeated.
Pause points seemed very reproducible in experiment from bead to bead
(each attached to a genetically identical DNA),
even when force and temperature varied considerably.
This statement will be quantified below.

In order to compare simulation results to experimental data, we converted
microns unzipped to the numbers of base pairs unzipped.  The centers of
beads attached to fully zipped strands of \lp{} DNA under a force near
15 pN were observed 16.5\mm{} from the round capillary in experiments.  The
centers of beads attached to fully unzipped strands of \lp{} DNA under a force
of 15 pN were observed 77.4\mm{} from the round capillary in experiments.
Thus DNA strands being unzipped were stretched to a length of $60.9\mm$.
Lambda phage DNA is $48,502$ base pairs long, so to convert from \mm{} to base pairs,
we use the conversion factor $48,502 \mathrm{bp}/60.9 \mm \approx 800 \mathrm{bp}/\mm$.
Since two strands of ssDNA are produced during unzipping, the monomer spacing along
a ssDNA strand
is found from the inverse of this factor divided by two to be $a\approx 0.6 nm$.
Such a linear interpolation seems reasonable given the fairly large forces
($\sim 15-20 pN$) acting on the unzipped `handles'.

\section{Pause Point Algorithm}
\label{sec:pause}

\begin{figure}
\vspace{0.2cm}
\includegraphics[scale=1]{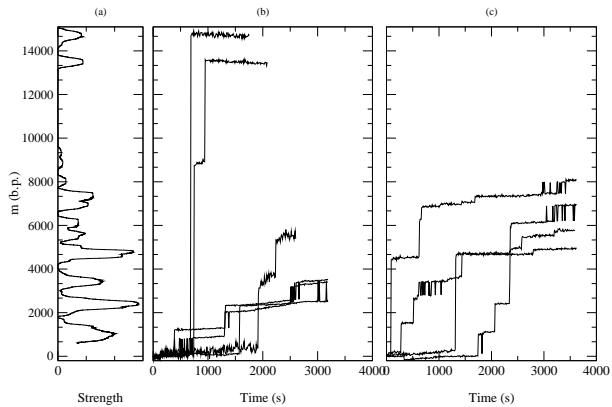}
\caption{\label{fig:spectrum_exp} (a) Experimental pause point spectrum
alongside sample experimental unzipping trajectories at $T = 298K$ \RT{} with
(b) $F = 15 $pN and (c) $F = 20 pN$.  
The experimental spatial resolution is $0.5\mm$ (about 400 bp), which
determined the size of the window used to calculate the spectrum. The
structure of some peaks indicate multiple underlying pause points that
are only partially resolved at this resolution. The full experimental
pause point spectrum was computed with 15 experimental trajectories, only
10 of which are shown.}
\end{figure}

Figure \ref{fig:spectrum_exp}(b-c) displays several experimental
unzipping trajectories.  As can be seen in the trajectories, the
unzipping fork progresses through long pauses at specific locations,
separated by rapid transitions between these pauses.  Moreover,
a sample of trajectories from identical DNA sequences display a uniformity
in the locations at which the DNA unzipping pauses.  Note also that pauses
at certain locations seem consistently longer than others. 
From these considerations, we are motivated
to develop a method for combining many trajectories to form a distribution
reflecting the location and relative strengths of pause points.

A pause point `spectrum' can be computed as follows (see Figure
\ref{fig:pause_examp} for an example):
\begin{figure}
\vspace{0.2cm}
\includegraphics[scale=1]{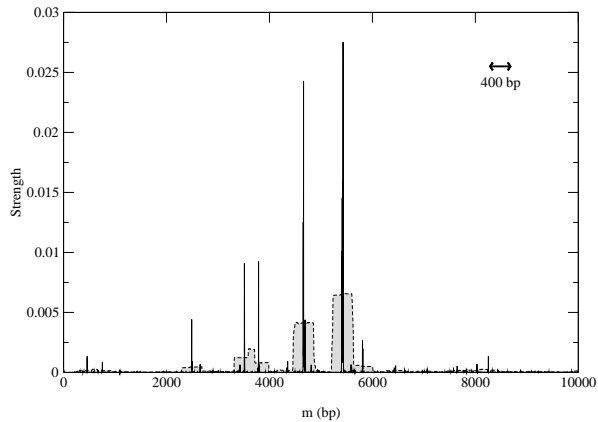}
\caption{\label{fig:pause_examp}Sample window averaged results.  
The high resolution (1 bp)
pause time histogram is shown in black.  The grey spectrum is created
by sliding a window of size 400 bp along the x-axis,
and assigning a y-value to the midpoint of the window equal to
the average of the high resolution histogram within that window.
Note that highly localized pause points appear as broad peaks 
according to a much lower resolution in the window average.}
\end{figure}
\begin{enumerate}
   \item Create a histogram (area normalized to 1) based on the position of the unzipping fork
         during the time duration of the experiment for all 
         trajectories using the highest resolution possible.
   \item In order to smooth this histogram according to the real experimental
         resolution, define a window centered around each position of the
         histogram, with a width equal to the experimental resolution.
   \item Compute the average histogram peak height within this window, and
         assign the value of this average to the position of the center of
         the window in the pause point spectrum.
\end{enumerate}
The location of the peaks in a pause point spectrum correspond to the distances
at which the trajectories paused, and thus correspond to pause points.
In addition, the peak area is proportional to the amount of time
trajectories spent at the peak locations. Hence, relative peak area can be used as
a measure of the relative strength of pause points. As can be seen from direct 
comparison between the
spectrum and trajectories (Figure \ref{fig:spectrum_exp}),
this method of analysis provides an excellent summary of pauses and jumps
observed in experiment.  From this intuition, we expect that higher
experimental spatial resolution would allow us to resolve further peaks in
the spectrum.


The experimental pause point spectrum was computed using 15 unzipping trajectories
of which 10 are shown in Figure \ref{fig:spectrum_exp}(b-c).
In addition to applying the above algorithm, the following steps were
taken:  Trajectories
were first individually shifted so that the starting position of
each trajectory was zero \mm.  This was done in order to subtract
the linker length from the measured distance for each bead.  In
order to collect as many trajectories as possible for better statistics,
the experimental resolution was 0.5\mm.
The spectrum was thus created with a window of 0.5\mm, 
corresponding to about 400 bp.
In each of the experimental trajectories, there was a period in
the beginning that was noisy due to transient bead adjustments in the 
turning on of the magnetic field 
(Figure \ref{fig:spectrum_exp}).
These regions were not included in the experimental spectra.
To convert from \mm{} to bp, we multiplied by (48502 bp/ 60.9 \mm),
which represents the appropriate bp/\mm{} factor for \lp{}
under these experimental conditions (Section \ref{sec:exp}).

\section{Theoretical Study of Pause Points}

\subsection{Defining the Free Energy Landscape}
\label{sec:theory:FEL}

We consider the DNA unzipping experiment (Figure \ref{fig:model}) as a chemical reaction from
dsDNA $\to$ 2 ssDNA.  This system possesses a natural one-dimensional reaction
coordinate, namely the number of base pairs (bp) unzipped $m$, or equivalently the spatial
location of the unzipping fork.  Theoretical descriptions can then be naturally reduced from
a complicated, three-dimensional system to a one-dimensional description with
some of the interactions renormalized, expressing the three-dimensional 
nature of the problem.

A very simple one-dimensional effective model 
\cite{Lubensky00,Lubensky02} can then be written down based on
the above picture.  We define a free energy as a function of the number of bases
unzipped, \e{m}, which represents the difference in free energy between the states
with $m$ base pairs unzipped and the fully zipped states ($\e{m=0}=0$).
There is a contribution to $\e{m} - \e{m-1}$ from the free energy difference
between the bound and unbound $m^{th}$ base pair, 
$\Delta G_{\mathrm{bp}} =  k_B T \widetilde{\eta}(m)$, 
as well as a contribution due to adding two additional monomers to the 
free ssDNA strands under a tension, $F$, denoted as $2 k_B T g(F)$.  
We can write these contributions as
\[
\label{eq:etatilde}
   \frac{\e{m} - \e{m-1}}{k_B T} = 2g(F) + \widetilde{\eta}(m).
\]
The DNA sequence information is stored in the function $\widetilde{\eta}(m)$.  
In principle, this function might depend on time
due to transient bubbles or twists in the DNA. At forces low
enough ($F \leq 5 pN$) such that hairpin formation in the unzipped handles
is possible \cite{Montanari01,Dessinges02}, $g(F)$ could also be sequence dependent.
These compilations are neglected here. 
If we iterate equation \eqref{eq:etatilde} until we reach $m=0$, we have
\[
\label{eq:eps}
\frac{\e{m}}{k_B T} = 2g(F)m + \sum_{n = 0}^m \widetilde{\eta}(n),
\]
Here we are setting $\widetilde{\eta}(0) = 0$, which ensures
$\e{0} = 0$.  The long DNA sequences we are considering should be
insensitive to such edge effects.

Thermodynamically we expect in equilibrium that unzipping of a dsDNA molecule of
$M$ bases will occur when $\e{M} < \e{0}$, with the transition region between ds and ss DNA
occurring when $\e{M} = \e{0}$.  If we define the shifted function
$\eta(n) = \widetilde{\eta}(n) - \overline{\eta}$, where $\overline{\eta}$
is the average of $\widetilde{\eta}(n)$ over the sequence, we can
rewrite \eqref{eq:eps} in the form
\[
\label{eq:fel}
   \frac{\e{m}}{k_B T} = fm + \sum_{n=0}^m \eta(n)
\]
\[
   f = 2g(F) + \overline{\eta}
\]
The parameter $f$ is a reduced-force, which defines the overall
tilt of the free energy landscape (\FEL{}), with the particular base sequence overlaid on
this tilt with the function $\sum_{n=0}^m \eta(n)$.  
With this definition, the critical reduced force is defined by the equation
$f = 0$.  Values of $f > 0$ represent forces too low to unzip, while values $f < 0$
represent forces where full equilibrium unzipping is thermodynamically
favorable.  In all of the above, we have
assumed that thermal bubbles do not form under the experimental
unzipping conditions.  This simple model can be defined in a more rigorous fashion by
integrating out three-dimensional degrees of freedom in a
statistical mechanical microscopic definition of the system. Effects due
to bubbles can be incorporated into a coarse grained model, with
renormalized parameters \cite{Lubensky02}.  

It is interesting to note that even for this simple model, 
non-trivial phenomena can occur due to the buildup of free energies
naturally present in \eqref{eq:fel}.  Indeed, for the case of a
completely random base sequence of length $M$, a sum over the independent
random variables in \eqref{eq:fel} would give an energy barrier  
$\sim k_B T\sqrt{M}$ \cite{Lubensky02}.
Since GC base pairs are $\sim \kB T$ stronger than AT pairs at
room temperature (Table \ref{tab:bq}),
 we expect large peaks to appear due to the presence of 
long GC-rich regions. For a sequence of length $M = 48,000$,
one expects barriers of the order of 200 $\kB T$.

In practice, \FEL's are computed for a particular genome
sequence using the experimentally determined free energies
of base quartet formation \cite{SantaLucia96}.  There are
10 distinct base quartets, where $m$ now represents the
$m^{\mathrm{th}}$ base quartet, while $\widetilde{\eta}(m)$
represents this base quartet's free energy (Table \ref{tab:bq}).  
\begin{table}
\begin{tabular}{cc}
\hline
Base Quartet & $\Delta G_{qt}/k_B T$ \\
\hline
5'-GC-3' &  4.46 \\
CG &  4.22 \\
GG &  3.46 \\
GA &  2.79 \\
GT &  2.96 \\
CA &  2.79 \\
CT &  2.20 \\
AA &  2.31 \\
AT &  1.52 \\
TA &  1.33 \\
\hline
\end{tabular}
\caption{\label{tab:bq} Base quartet free energies $\Delta G_{qt}$ for
the bound to unbound transition
for $T = 298 K$ \RT{} taken from \cite{SantaLucia96}, using
$\Delta G_{qt} = \Delta H_{qt} - T\Delta S_{qt}$. 
Only two nucleotides of the base quartet are shown, the other
two obtained from the usual complementarity A-T and G-C.
Free energies are expressed in units of $k_B T = 0.59\mathrm{kCal/mol}$
at $T = 298K$ \RT.}
\end{table}
These free energy
parameters are determined through thermal denaturation studies
on short dsDNA fragments, and were found for temperatures of
around 310 $K$.  By using base
quartet free energies, base stacking interactions are included,
which are thought to be more important for overall dsDNA
stability than the hydrogen bonds in between base pairs 
\cite{Grosberg94,Blossey03}.
To compute \FEL's for different temperatures,
the free energies for a given quartet were calculated from
$k_B T \widetilde{\eta}(m) = \Delta G_{qt} = \Delta H_{qt} - T\Delta S_{qt}$.  
Here $\Delta H_{qt}$ ($\Delta S_{qt}$) is the enthalpy (entropy)
difference between the bound and unbound DNA base quartet.
Once temperature
and $f$ are specified, the \FEL{} is computed with equation
\eqref{eq:etatilde} or \eqref{eq:fel}.

\begin{figure}
\vspace{0.6cm}
\includegraphics[scale=1]{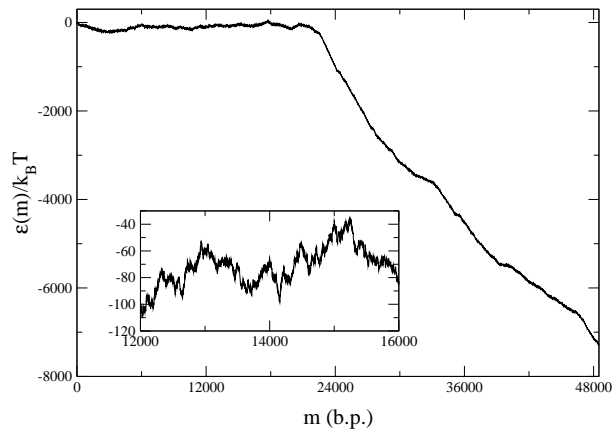}
\caption{\label{fig:lambda_FEL} Free energy landscape for the \lp{} genome at $f = -0.15,
T = 298K$ \RT{} corresponding to $F \approx 16 $pN.  A closer view is given in the inset.
Because the lambda phage genome splits into a GC-rich `front end', followed
by an AT-rich region, the effective critical unzipping force is the one shown
here, which produces an approximately flat energy landscape for the first
$\sim 20,000$ base pairs.  Note that there are still energy barriers
$\sim 20 k_B T$ in this region due to sequence heterogeneity.}
\end{figure}

The case of \lp{} DNA is particularly interesting since it is
known that this genome consists of a GC-rich half connected
to an AT-rich half \footnote{The lambda phage genome can
be found at http://www.ncbi.nih.gov/ with sequence
accession number NC\_001416.}.  Using the free energy 
parameters of \cite{SantaLucia96}, one finds that the large
GC-rich region creates a peak of approximately $3000 \kB T$
at $f=0$ and $T=298K$ \RT{}, representing an insurmountable barrier
to unzipping, and which is much larger than that expected for
a random sequence.  For the \lp{} genome, we thus define an operational
critical reduced force of unzipping as the value of $f$ such that 
$\e{0} = \e{m_{\textrm{GC}}}$, where $m_{\textrm{GC}}$ is
the boundary of the GC-rich region. 
As can be seen in Figure
\ref{fig:lambda_FEL}, this operational critical reduced force corresponds to
approximately $f = -0.15$.
Forces greater than this should allow easier unzipping since the AT-rich
portion of the \FEL{} has a negative slope for these forces.
However, even at these large forces, there are barriers on
the order of 20$\kB T$ to unzip (Figure \ref{fig:lambda_FEL} inset.)
Using a Freely-Jointed Chain (FJC) model for the single 
strands, with monomer spacing, $a = 0.6$nm (corresponding to
the bp/\mm{} conversion factor in section \ref{sec:exp}), and Kuhn length, 
$b = 1.9$nm \cite{Smith96}, a reduced force value of $f = -0.15$ 
corresponds to an experimental force value of $F \approx 16 $pN.

\subsection{Dynamics}
\label{sec:theory:MC}

To study the location of the pause points one has to study the
dynamics related to moving along the chemical reaction coordinate, $m$.
Macroscopic unzipping occurs only if $f<0$, when the equilibrium state of the
system is unbound.
In this case, the experimental traces of DNA unzipping represent the approach
of the system toward its equilibrium single-stranded state.

As outlined in \cite{Lubensky02}, there are four dynamical
time scales associated with DNA constant-force unzipping in the
setup shown in Figure \ref{fig:model}: \ts{end} and \ts{bulk}
represent base pairing and unpairing at the end of the strand
and in the bulk, respectively; $\ts{ss}(m)$ represents the relaxation
time of the liberated single strands; and $\ts{rot}(m)$ represents
the relaxation time of twist built up in the zipped portion
of the strand due to the helical nature of the DNA.  The latter
two time scales vary as a function of $m$. The dynamics
of unzipping are determined by the slowest of these time scales.
Here we assume that this time scale, for any value of $m$,
is related to the unbinding of base pairs.

In the analysis below, we assume that the slowest timescale is
$m$-independent.  Furthermore, it can be argued that bubble
formation is suppressed in DNA for the relevant experimental conditions 
because of strong base stacking interactions \cite{Blossey03}.

We will be interested in the unzipping dynamics for $f < 0$, that
is for forces above the critical force of unzipping where it is
thermodynamically favorable to unzip.  However, even under these
conditions, the approach to thermodynamic equilibrium is far from
simple. Smooth progress of the unzipping fork is hampered by
very large energy barriers in \e{m} that can be caused by the
buildup of positive $\eta(m)$.  As mentioned above, for random
DNA sequences of length $M$, these barriers can grow as
$\sqrt{M}$.  Forces slightly above the critical force are unable
to remove these barriers through tilting the landscape, and we
expect to observe difficulty in traversing these barriers.  
Since the barriers are sequence dependent, it is possible that
the dynamics of
unzipping display characteristic signatures of the sequence.

To study the behavior of a particular DNA sequence,
and to make direct contact with experiments, it is 
useful to have a dynamical model that closely mimics
the experiment.  This can be achieved
most simply through Monte Carlo (MC) simulations of a random
walker on the one-dimensional \FEL{} for the specific
DNA sequence under study, at the specified reduced force and
temperature conditions.  The position of the walker on
the \FEL{} represents the position of the unzipping fork
in experiments.  The walker moves from position $m$ to a
nearest neighbor position $m\pm 1$ with the rate
\[
 w[m\to (m\pm 1)] = \min \left\{1,e^{-[\e{m\pm 1} - \e{m}]/\kB T}\right\}.
\]
Details of the algorithm are outlined in Appendix \ref{app:MC}.

\begin{figure}
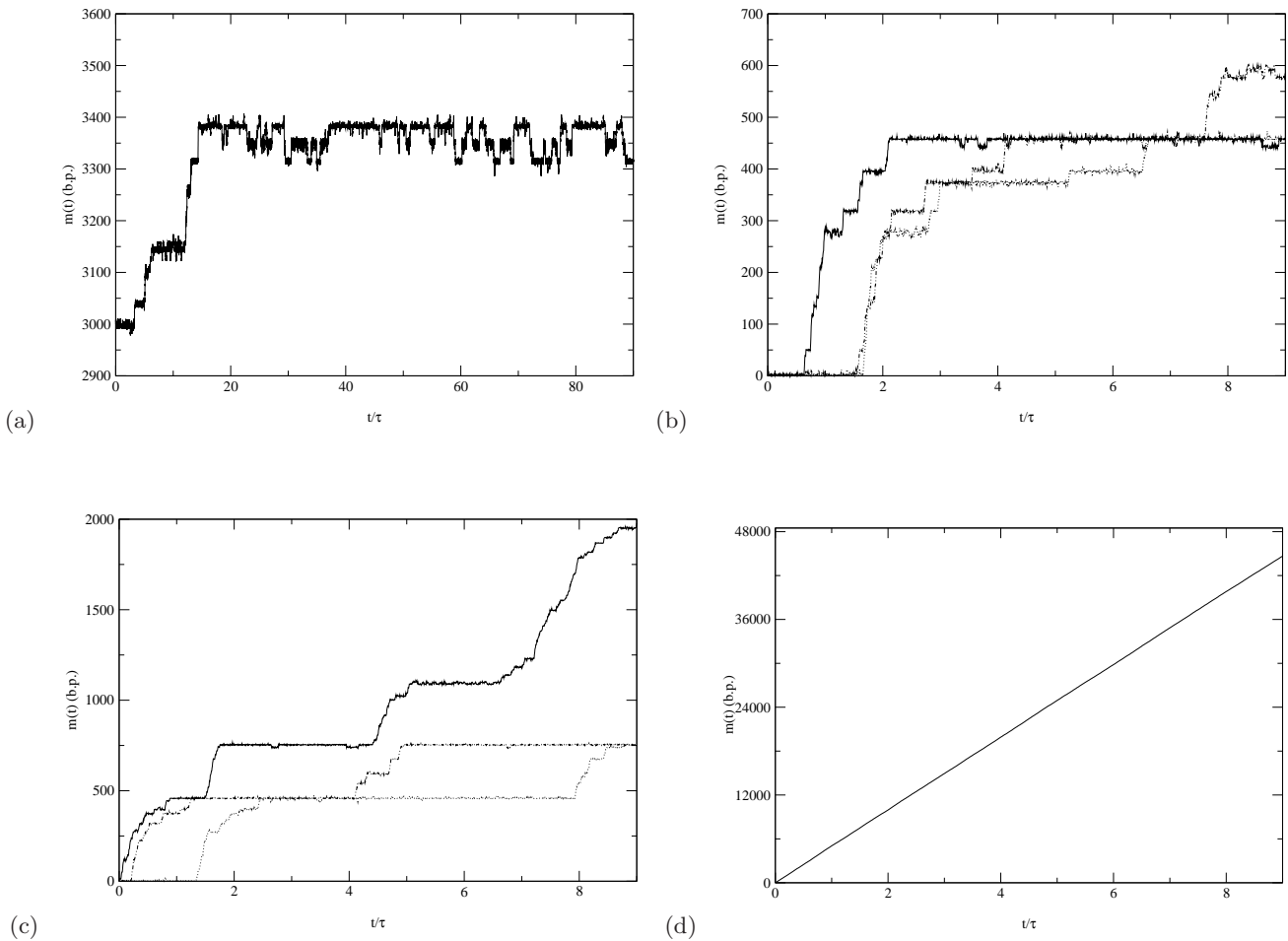

\begin{tabular}{cc}
(a) \hspace{0.1cm} \includegraphics[scale=1]{sample_trajs_fn0p26.eps} \vspace{0.5cm} &
(b) \hspace{0.1cm} \includegraphics[scale=1]{sample_trajs_fn0p39.eps} \vspace{0.5cm} \\
(c) \hspace{0.1cm} \includegraphics[scale=1]{sample_trajs_fn0p50.eps} &
(d) \hspace{0.1cm} \includegraphics[scale=1]{sample_trajs_fn5.eps} \\
\end{tabular}
\caption{\label{fig:theory_traj} Sample theoretical unzipping
trajectories, $m(t)$, where $\tau = 10000$ steps.  (a) $f = -0.26$, 
(b)$f = -0.39$, (c) $f = -0.50$, (d) $f = -5$.  For
$t/\tau \geq 15$, (a) shows intricate
two-state behavior caused by nearly degenerate minima on the \FEL.
In (d) the unzipping is smooth,
but does not fully unzip in $80,000$ time steps indicating dwell
time at some sites.  Note that pause points are reproducible
in the simulations, and that very large forces are
required to smooth out the large barriers present in \lp{} DNA
and allow smooth unzipping.}
\end{figure}

A simulation consists of specifying the \FEL{} (DNA sequence,
temperature and reduced force), and propagating the MC algorithm
for a specified number of steps.  The initial condition is such
that the walker starts at $m = 0$, representing the experimental
circumstance of tracking DNA's that begin as fully zipped.
What results is trajectory
data, $m(t)$, which contains the same information obtained in experiments.
Sample theoretical trajectories are shown in 
Figure \ref{fig:theory_traj}.  There are clear pauses and jumps 
of the trajectories for reduced forces much higher than the critical
force ($ -0.39 \leq f \leq -0.5$).  Only extremely large reduced forces 
($f = -5 $) are sufficient to remove all barriers and 
allow smooth unzipping.

It must be stressed that there is a certain freedom in 
choosing particular MC algorithms (Appendix \ref{app:MC}).
Algorithms can differ in how long it takes random walkers
to traverse energy barriers, and thus we do not expect
to be able to compare timescale information between experiment
and theory, quantitatively.  Although pause point locations
can be predicted, we do not expect
pause point strengths to match between experiment and theory.

To calculate the theoretical pause point spectra,
simulations were performed at a variety of reduced forces, $f$,
and on \FEL's at the same temperature of the experiments.
Each simulated trajectory consisted of $10^{7}$ Monte Carlo
steps, and 300 trajectories were used to create the theoretical
histograms.  Corresponding to the 0.5\mm{} experimental resolution, the
window averages were taken over 400 base pairs with a length
per base pair for \lp{} DNA under these conditions of
60.9\mm/48502 bp as discussed in Section \ref{sec:exp}.
For $f$ values
in which some of the trajectories reached the fully unzipped
state ($m = 48,502$ bp), the trajectories were cutoff past $48,000$ base
pairs before being included in the spectra.

\section{Discussion}
\label{sec:disc}

An examination of a few experimental unzipping trajectories 
(Figure \ref{fig:spectrum_exp}) reveals
that the pausing locations are often encountered by multiple copies
of the same DNA.  These copies are subject to different realizations
of thermal noise, but share the same base sequence, an indication
that sequence is a strong factor in governing pause point locations.

Experimental and theoretical pause point locations can be compared
by examining the peak positions in the corresponding pause point
spectra (Table \ref{tab:peak_loc}).
\begin{table}
\begin{tabular}{cc}
\hline
Experiment ($\pm 200$ bp) & Theory ($\pm 200$ bp)\\
\hline
1000 &  600 \\
2400 &  2500 \\
3400 &  3400 \\
4700 &  4600 \\
5600 &  5400 \\
6100 &  \\
7100 & \\
Gap = 6400 & Gap = 8800 \\
13500 &  \\
      &  14200 \\
14700 & \\
\end{tabular}
\caption{\label{tab:peak_loc} Experimental vs. theoretical
pause point locations (bp) for the first 15000 bp corresponding
to unzipping the front-half of the lambda phage genome
(Figure \ref{fig:lambda_FEL}).  The pause point locations are
the positions of the centers of the peaks in the pause point
spectra (Figure \ref{fig:exp_vs_theory}), and have errorbars
of $\pm$ 200 bp due to experimental resolution.  The theoretical pause points
include those found for $f = -0.29, -0.39, -0.47, -0.50$.  Also listed
are the size of the gap regions in the spectra where no pause points are 
found.  Note that every theoretical peak less than 6000 bp is within the errorbars of
an experimental peak, and the theoretical and experimental gaps
are roughly the same size and in the same location in the pause
point spectra.}
\end{table}
There is strong agreement between experimental and theoretical
pause point locations at distances less than 6000 bp.  In addition,
theory predicts a gap in the pause point spectra of $\sim 9000$ bp
starting at 5400 bp,
which is similar in size and location to that observed in experiment.
The fact that the positions of the pause points and gaps match to such a high degree between
experiment and theory are evidence that for these experimental conditions,
the approximations inherent in the concept of dynamics on the \FEL{}
representing DNA constant-force unzipping as a model for the experiments
are sound.  Since the theoretical model only requires the base
sequence and thermodynamic parameters for base quartet formation,
the agreement is proof that pause point locations are strongly
governed by base sequence. The good comparison also shows that
neglect of bubble formation for these
temperatures is appropriate, as is also found by other means 
\cite{Blossey03}.

\begin{figure}
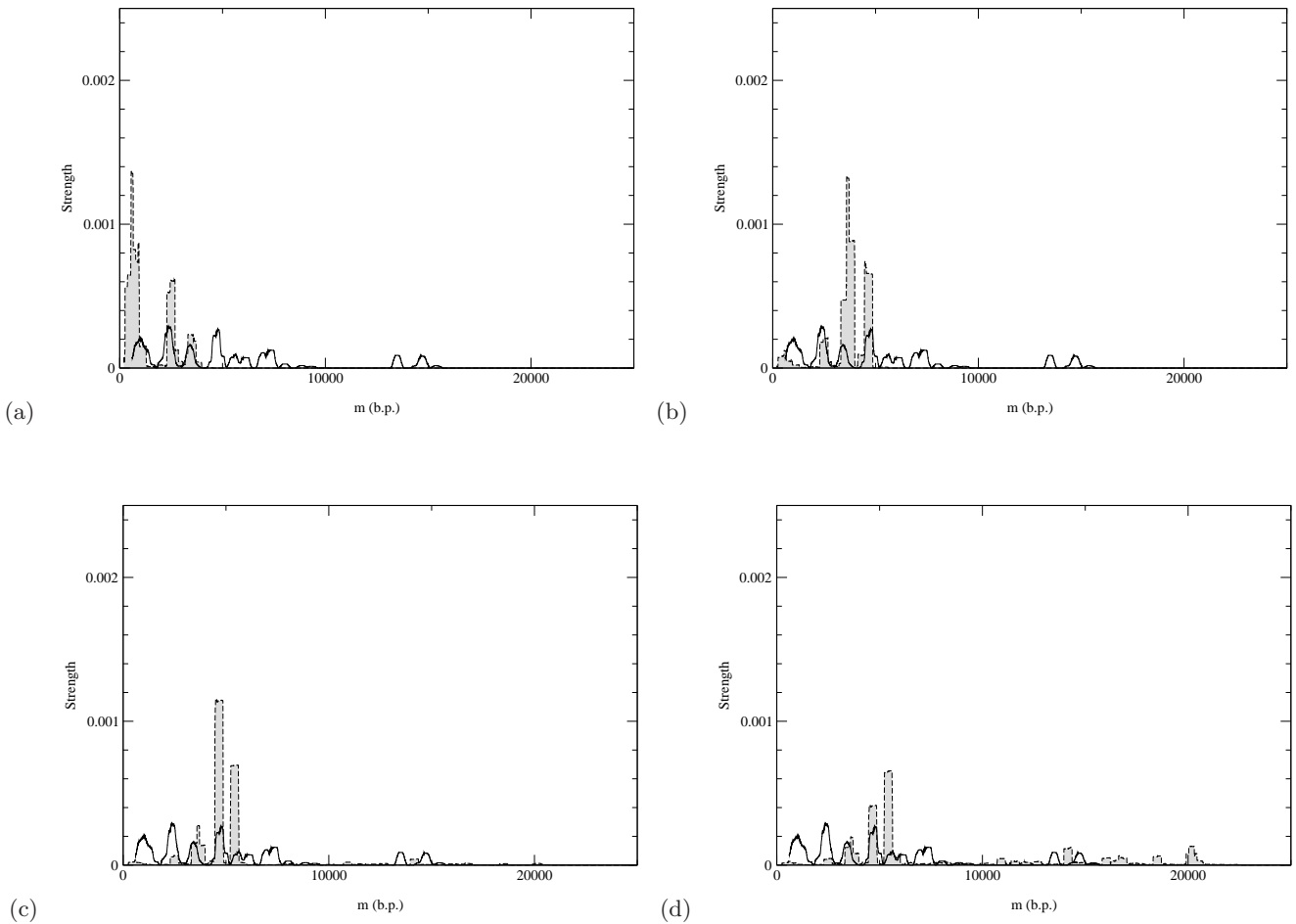

\begin{tabular}{cc}
(a) \hspace{0.1cm} \includegraphics[scale=1]{exp_comp_bp_RT_fn0p29.eps} \vspace{0.5cm} &
(b) \hspace{0.1cm} \includegraphics[scale=1]{exp_comp_bp_RT_fn0p39.eps} \vspace{0.5cm} \\
(c) \hspace{0.1cm} \includegraphics[scale=1]{exp_comp_bp_RT_fn0p47.eps} &
(d) \hspace{0.1cm} \includegraphics[scale=1]{exp_comp_bp_RT_fn0p50.eps} \\
\end{tabular}
\caption{\label{fig:exp_vs_theory} Experiment (black) and theoretical
(grey) pause point spectra for \lp{} at $T = 298K$ \RT{}. 
Section \ref{sec:pause} outlines how the spectra were created using
a sliding window average based on a 0.5\mm{} experimental 
resolution.  Experiments were done with
$F = 15 $pN and $F = 20$ pN, 
and the spectrum was obtained from 15 experimental
traces (see Figure \ref{fig:spectrum_exp}).  
Theoretical spectra are at (a) $f = -0.29$, (b) $f = -0.39$,
(c) $f = -0.47$, (d) $f = -0.50$, and where created using 300
traces of $10^7$ steps each.}
\end{figure}

The pause point spectra contain much more information than the
pause point locations.  Theoretical and experimental pause point
spectra are shown in Figure \ref{fig:exp_vs_theory}.
The values of $f$ used in the simulations lie
in the range $-0.25 \leq f \leq -0.5$. Recall that $f = -0.15$ corresponds
roughly to a flat average \FEL{} in the GC-rich region of the \lp{}
DNA (Figure \ref{fig:lambda_FEL}). A value of $F \approx 17 $pN corresponds to $f = -0.37$ under these
conditions, which is within this range. We have compared experimental and
theoretical pause points in the \emph{front} half of the unzipping process.  The
much steeper energy landscape in the back half (see Figure \ref{fig:lambda_FEL}) eliminates
most pause points.
As the values of $f$ are gradually decreased from $f= -0.25$, the
theoretical pause point spectra grow into more peaks at larger distances,
 although low base pair peaks are still preserved.  Thus
the locations of pause points are fairly robust with respect
to $f$ values.
Once the forces are high enough to allow exploration of the whole \FEL{},
the location of the peaks in the spectra do not change, and peak areas
are adjusted reflecting a changing of the strength of the pause points.

There are some noticeable disagreements in pause point location
between experiment and theory.  In particular, the experimental
doublet peak at 7100 bp was not observed in any theoretical spectra for
a variety of parameters, including longer simulation times.  The pair
of peaks centered around 14000 bp in the experimental spectra are also
not picked up in the theoretical spectra, rather a single peak lying
in the middle of the experimental peaks is found.  This most likely
does not represent an averaging of the two pause point locations in
the theoretical spectrum because we would expect a broad peak covering
the two locations in this scenario.  The doublet of experimental peaks
represents data from two separate runs, and thus a slight miscalibration
in the $800 bp/\mm$ conversion factor (Section \ref{sec:exp}) 
for those particular runs could cause the two peaks to separate, since
a miscalibration has a larger effect for longer distance pause points.
For strong enough forces, theoretical
pause point spectra also display many more peaks than present in the
experimental spectra.  It could be that more experimental trajectories
need to be included to observe these peaks.

While the peak positions in the experimental and theoretical pause point
spectra coincide quite well, the peak areas noticeably differ. One source
of this discrepancy is due to the time scales of the DNA unzipping.
Each step in the Monte Carlo propagation of the unzipping can be thought
to occur on the microscopic time scale governing the DNA unzipping, which
is estimated to be $\sim 10^{-7}$s \cite{Danilowicz03a}, with a large
error in the exponent \cite{Mathe04}.  The accuracy of this figure is not high enough
to allow direct comparison with theoretical and experimental time scales.
As mentioned above, the particular choice of Monte Carlo algorithm can
change the characteristic unzipping times of simulations and could account
for the discrepancy between theoretical and experimental time scales.

\begin{figure}
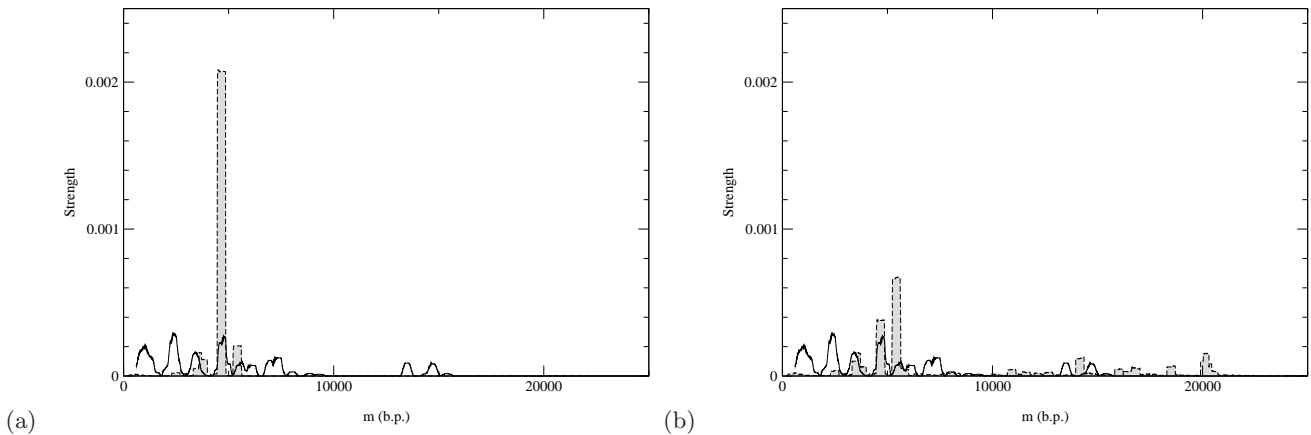

\begin{tabular}{cc}
(a) \hspace{0.1cm} \includegraphics[scale=1]{exp_comp_bp_RT_fn0p39_long.eps} &
(b) \hspace{0.1cm} \includegraphics[scale=1]{exp_comp_bp_RT_fn0p50_long.eps} \\
\end{tabular}
\caption{\label{fig:spectra_long}Experimental (black) and 
theoretical (grey) pause point spectra for \lp{} simulations 
of $10^8$ steps at $298K$ \RT{}. (a) $f = -0.39$, (b) $f = -0.50$.  
The longer run times in the simulations produce different
pause point strengths, but similar pause point locations
to Figure \ref{fig:exp_vs_theory}.}
\end{figure}

Figure \ref{fig:spectra_long} shows pause point spectra obtained for simulations
of $10^8$ steps.
A reduced force of $f = -0.39$ is not strong enough to allow DNA's to unzip
under this length of time.  Comparing to the $10^7$ step simulations
(Figure \ref{fig:exp_vs_theory}), we see that longer times in this case allow peaks
at slightly higher base pair to be observed, but mainly result in a
change in peak area.  For $f = -0.50$, which allows for some fraction 
of unzipping even with $10^7$ step runs, longer runs only serve to change
peak areas (compare with Figure \ref{fig:exp_vs_theory}).  Longer simulation times do
not give spectra that approach the experimental pause point
spectrum.  We thus expect that even longer times will not provide quantitative
agreement of pause point strength with the current MC algorithm.
The choice of the Metropolis MC algorithm is an efficient choice
to satisfy the detailed balance condition \eqref{DB}, but it is not
the only choice for MC algorithm.  Other choices for algorithms can
give different pause times at pause points which result in different
peak areas in pause point spectra.

The discrepancy of unzipping timescales between experiment and theory
is also reflected in the fact that multiple values of $f$ needed
to be used to obtain the theoretical pause point information. 
At low $f$ values, theoretical simulations could not pass certain
barriers of the \FEL{}, as is seen in the abrupt cutoff of peaks
in Figure \ref{fig:exp_vs_theory}(a-c), which necessitated further
tilting of the landscape.  The technique of increasing the force
is also used in experiment to probe farther out regions of the 
unzipping landscape (Figure \ref{fig:spectrum_exp}).  However, the range
of reduced forces used in the theoretical simulations corresponds roughly
to 15-17pN, while the experimental range is roughly 15-30pN 
\cite{Danilowicz03b}.  

We might expect discrepancies between experimental and theoretical
pause point locations to be due to large A-rich regions in the
genome, since these are more susceptible to bubble formation due
to the weaker base pairing and stacking interactions (Table \ref{tab:bq}).
Thus further investigation into these pause point discrepancies can
lead to interesting genomic information.  Experiments involving
higher temperatures and different ionic conditions will help 
elucidate these discrepancies \cite{Danilowicz03b}.

\begin{figure}
\vspace{0.5cm}
\includegraphics[scale=1]{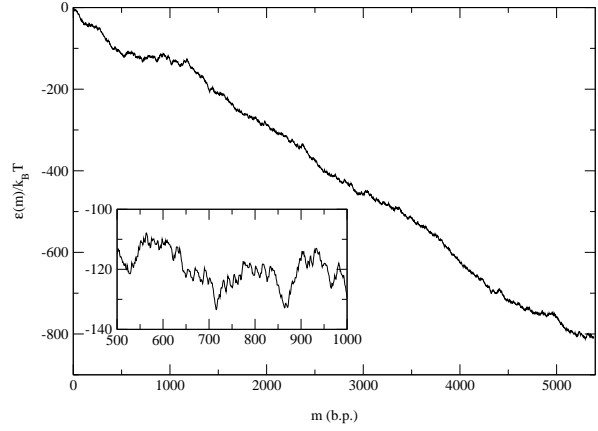}
\caption{\label{fig:bphi_FEL} Free energy landscape for the \bphi{} genome at $f = -0.15,
T = 298K$ \RT{} corresponding to $F \approx 16 $pN.  A closer view is given in the inset
of the approximately horizontal plateau region.}
\end{figure}

\begin{figure}
\vspace{0.5cm}
\includegraphics[scale=1]{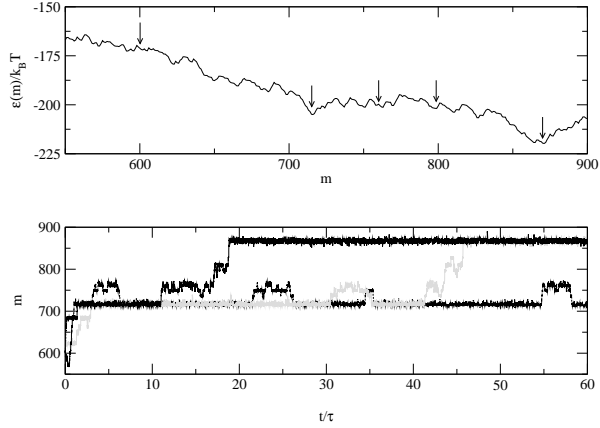}
\caption{\label{fig:bp_traj} Sample simulation trajectories displayed
below the relevant segment of the \bphi{} \FEL{} at $f = -0.25$ \RT.  The
three trajectories start at the left of the figure at 
$m = 600 \mathrm{ bp}, t= 0$.  Simulation time has units $\tau = 5385$ steps corresponding
to the genome length. Pause points are denoted by arrows on the \FEL.  Note that
very intricate multi-state behavior is seen in the walker trajectories.  In particular,
the region from $725-775$ bp shows the presence of several minima of the same depth
on the \FEL, and shows up as oscillations in the trajectories.}
\end{figure}

\begin{figure}
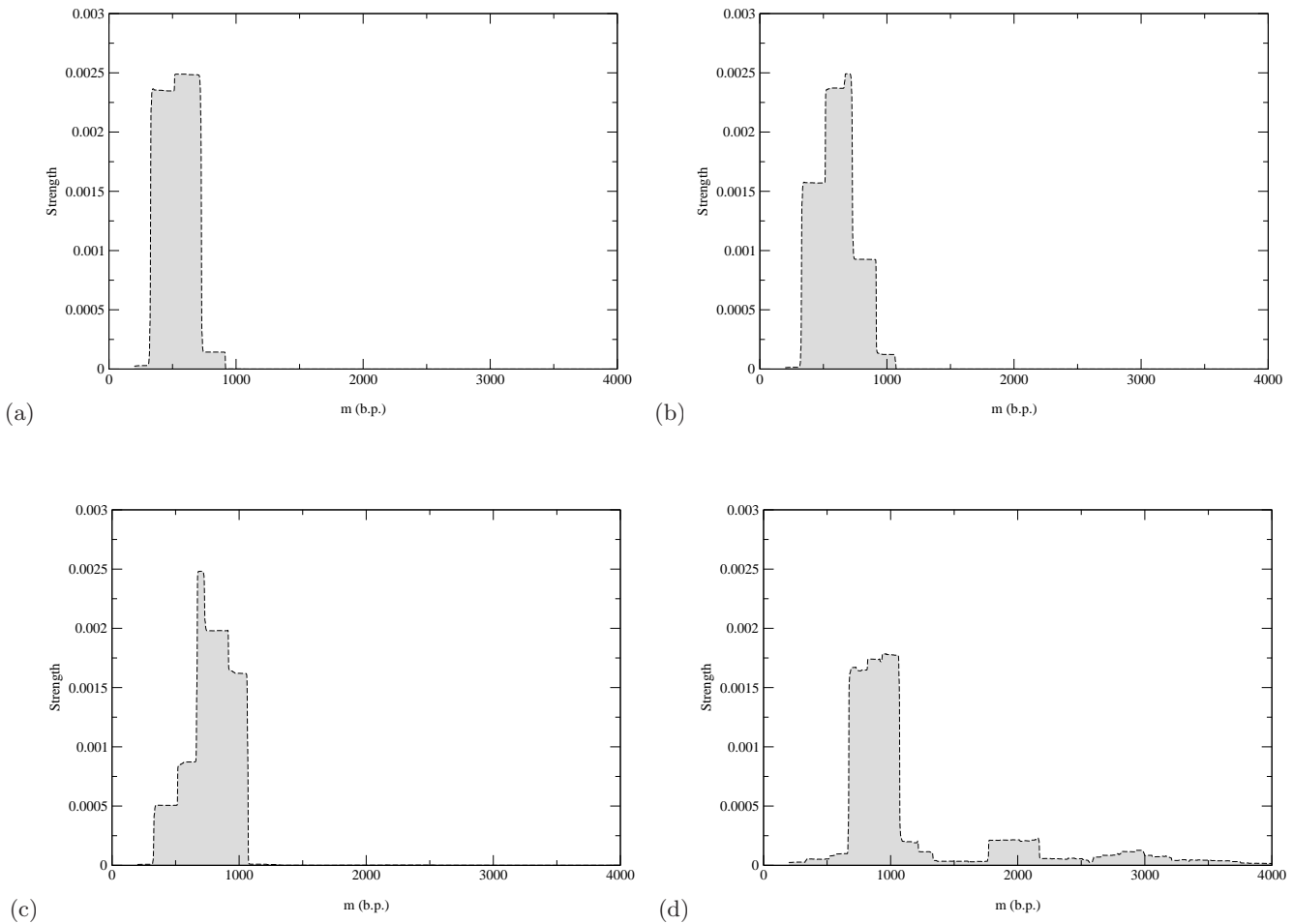

\begin{tabular}{cc}
(a) \includegraphics[scale=1]{bp_phiX174_RT_fn0p15.eps} \vspace{0.5cm} &
(b) \includegraphics[scale=1]{bp_phiX174_RT_fn0p20.eps} \vspace{0.5cm} \\
(c) \includegraphics[scale=1]{bp_phiX174_RT_fn0p25.eps} &
(d) \includegraphics[scale=1]{bp_phiX174_RT_fn0p45.eps} \\
\end{tabular}
\caption{\label{fig:bp_spectra} Theoretical pause point spectra for
\bphi{} at $T = 298K$ \RT{}: (a) $f = -0.15$, (b) $f = -0.20$,
(c) $f = -0.25$, (d) $f = -0.45$. Each spectrum was created using 300
traces of $10^7$ steps each. A value of $f = -0.45$ corresponds to
$F \approx 17 $pN.}
\end{figure}

This procedure to investigate pause points in DNA unzipping is easily
extended to the study of other genomes since all that is required is
the base sequence and temperature of interest.  As an example, we theoretically
investigated the pause point unzipping spectrum for the microvirus
Bacteriophage Phi-X174 (\bphi{}) ($M = 5386$) \footnote{The \bphi{} genome can
be found at http://www.ncbi.nih.gov/ with sequence
accession number NC\_001422.}.  The \FEL{} for
\bphi{} has barriers that are on the order of $\sqrt{M}$, and is a
good example of a landscape which can be approximated by an integrated
random walk (Figure \ref{fig:bphi_FEL}).  
Figure \ref{fig:bp_traj} plots several sample simulation trajectories
alongside a segment of the \bphi{} \FEL{} for $f = -0.25$, and
figure \ref{fig:bp_spectra} plots pause point 
spectra for several values of $f$, all at $T = 298K$ \RT.  Once again
we see that for low values of $f$, the spectra grow into peaks at
higher base pair as the value of $f$ is increased.  A value of $f = -0.45$
is large enough to cause unzipping, and we can see that the spectrum
at this value has contributions from the whole surface.  For \bphi{},
$f = -0.45$ corresponds to $F \approx 17 $pN at $298K$ \RT.

\section{Conclusion}

We have presented experimental evidence that the
dynamics of DNA constant force unzipping are not smooth,
but rather display characteristic pauses and jumps.
Furthermore, we have given strong evidence that the locations
of these pauses are primarily governed by the DNA sequence
at room temperature \RT{}.
We have also presented a general scheme for computing
pause point spectra for any DNA sequence, with the only
inputs being the sequence, temperature, and a set of
ten empirical parameters representing DNA duplex stability.

The ideas presented above can be applied to any system in which the
concept of a pause point can be well defined, or in which a `spectrum'
representation of trajectory data can be useful in other ways.  We
can then enumerate the steps involved in constructing a theoretical
representation of the system in order to facilitate comparison with
experiments:
\begin{enumerate}
  \item Using chemical intuition, reduce the system to one degree of
        freedom.  Equilibrium statistical mechanics can be used to
        justify, or derive, the resulting \FEL{} description of the
        system.
  \item To model experiments, use Monte Carlo simulation with the
        appropriate algorithm to create theoretical trajectories.
  \item Compute trajectory spectra using the above procedure for
        both experimental and theoretical trajectories.
\end{enumerate}
Such systems, of which DNA constant
force unzipping is one, also include topics of current interest such
as the motion of molecular motors on biopolymers 
\cite{Keller00,Davenport00,Wang98,Neuman03,Perkins03}.

\section{Acknowledgments}

JBL would like to acknowledge the financial support of
the John and Fannie Hertz Foundation. Work by JDW, CD and
MP was funded by grants: MURI: Dept. of the Navy
N00014-01-1-0782; Materials Research Science and Engineering Center
(MRSEC): NSF Grant No. DMR-0213805 and NSF Award PHY-9876929. 
Work by DRN and YK was supported primarily by the National Science 
Foundation through the Harvard Materials Research
Science and Engineering Laboratory via Grant No.
DMR-0213805 and through Grant No. DMR-0231631.  YK
was also supported through NSF Grant No. 
DMR-0229243.

\appendix
\section{Monte Carlo Algorithm}
\label{app:MC}

The Monte Carlo technique is designed to sample an ergodic
system according to the equilibrium distribution
for long simulation times.  The distribution is specified
by the detailed balance condition
\[
\label{DB}
   \frac{w_{m \to m+1}}{w_{m+1 \to m}} = e^{-(\e{m+1}-\e{m})/\kB T},
\]
where $w_{m \to m+1}$ is the rate of taking the step
from $m$ to $m+1$ base pairs unzipped. The ratio on the right
hand side of \eqref{DB} insures relaxation to the Boltzmann
distribution for long times \cite{Newman99}.
Specifying the distribution, and thus the detailed balance
criterion, still offers a large degree of flexibility in
choosing an algorithm.  Our goal in this study is to be
able to predict the pause points of the DNA unzipping process,
and to this end, we expect many choices of Monte Carlo
algorithm to give equivalent pause points.  The simplest
algorithm to achieve the detailed balance is known as
the Metropolis Criterion \cite{Newman99}.  For an unzipping fork
location at $m$,
\begin{enumerate}
   \item Choose a direction to move ($m + \delta, \delta = \pm 1$).
   \item If $\e{m+\delta} - \e{m} < 0$, accept the move and \verb=GOTO= 1.
   \item If $\e{m+\delta} - \e{m} > 0$, accept the move with the probability
      according to the Boltzmann distribution
      ($e^{-(\e{m+\delta} - \e{m})/\kB T}$). Else stay at this $m$.
      \verb=GOTO= 1.
\end{enumerate}

In order to prevent the random walkers from trying to unzip
(rezip) beyond the end (beginning) of the dsDNA strand, we
artificially inserted infinite barriers to these transitions
in the simulations.


\end{document}